\documentclass[12pt,preprint]{aastex}

\slugcomment{Not to appear in Nonlearned J., 45.}

\shorttitle{Modeling of the zodiacal emission for the {\it AKARI} MIR maps}
\shortauthors{Kondo et al.}

\begin{document}


\title{Modeling of the zodiacal emission for the {\it AKARI}/IRC mid-infrared all-sky diffuse maps}


\author{Toru Kondo\altaffilmark{1}, Daisuke Ishihara\altaffilmark{1}, Hidehiro Kaneda\altaffilmark{1}, Keichiro Nakamichi\altaffilmark{1}, Sachi Takaba\altaffilmark{1}, and Hiroshi Kobayashi\altaffilmark{1}}
\affil{Graduate School of Science, Nagoya University}
\email{kondo@u.phys.nagoya-u.ac.jp}

\author{Takafumi Ootsubo\altaffilmark{2}}
\affil{Graduate School of Arts and Sciences, The University of Tokyo}

\author{Jeonghyun Pyo\altaffilmark{3}}
\affil{Korea Astronomy and Space Science Institute}

\and

\author{Takashi Onaka\altaffilmark{4}}
\affil{Graduate School of Science, The University of Tokyo}

\altaffiltext{1}{Graduate School of Science, Nagoya University, Chikusa-ku, Nagoya 464-8602, Japan}
\altaffiltext{2}{Graduate School of Arts and Sciences, The University of Tokyo, Meguro-ku, Tokyo 153-8902, Japan}
\altaffiltext{3}{Korea Astronomy and Space Science Institute, Daejeon 305-348, Republic of Korea}
\altaffiltext{4}{Graduate School of Science, The University of Tokyo, Bunkyo-ku, Tokyo 113-0033, Japan}


\begin{abstract}
The zodiacal emission, which is the thermal infrared (IR) emission from the interplanetary dust (IPD) in our Solar System, has been studied for a long time. Nevertheless, accurate modeling of the zodiacal emission has not been successful to reproduce the all-sky spatial distribution of the zodiacal emission, especially in the mid-IR where the zodiacal emission peaks. We therefore aim to improve the IPD cloud model based on \citet{kelsall98}, using the {\it AKARI} 9 and 18 $\micron$ all-sky diffuse maps. By adopting a new fitting method based on the total brightness, we have succeeded in reducing the residual levels after subtraction of the zodiacal emission from the {\it AKARI} data and thus in improving the modeling of the zodiacal emission. Comparing the {\it AKARI} and the {\it COBE} data, we confirm that the changes from the previous model to our new model are mostly due to model improvements, but not temporal variations between the {\it AKARI} and the {\it COBE} epoch, except for the position of the Earth-trailing blob. Our results suggest that the size of the smooth cloud, a dominant component in the model, is by about 10\% more compact than previously thought, and that the dust sizes are not large enough to emit blackbody radiation in the mid-IR. Furthermore we significantly detect an isotropically-distributed IPD component, owing to accurate baseline measurement with {\it AKARI}.
\end{abstract}





\section{Introduction}
\label{intro}

The zodiacal emission, which is the thermal infrared (IR) emission from interplanetary dust (IPD) in our Solar System, is a dominant foreground component in the mid-IR. \citet{cassini1685} studied the origin of the zodiacal light, sunlight scattered by the IPD grains, for the first time by visual observations of night sky. He considered that the IPD cloud has a lenticular structure centered on the Sun with its main axis lying on the ecliptic plane. In the early 1940s, Fessenkov considered the IPD distribution as a prolate spheroid surrounded by a dust torus, which is formed from fragmentation of asteroids in the asteroid belt \citep{struve43}. In the 1980s, IR observations with satellites, such as {\it IRAS} and {\it COBE}, revealed that the cloud has a more complicated structure. \citet{kelsall98} and \citet{wright98} constructed IPD cloud models, using the ten photometric band data of the {\it COBE}/Diffuse Infrared Background Experiment (DIRBE) from 1.25 to 240 $\mu$m. In addition to a smooth cloud arising from a mixture of dust associated with asteroidal and cometary debris, the models include asteroidal dust bands and a mean motion resonance (MMR) component trapped by the Earth into resonant orbits near 1 AU. In this study, we discuss the IPD cloud based on the model constructed by \citet{kelsall98}. \citet{kelsall98} removed the zodiacal emission from the DIRBE maps using their model (hereafter called the Kelsall model). However, there is a significant residual component of the zodiacal emission in the DIRBE mid-IR maps. For example, the brightness levels are $\sim$1 and $\sim$2 MJy~sr$^{-1}$ at the ecliptic poles and on the ecliptic plane, respectively, at 25 $\micron$ as shown in Figure 2(c) in \citet{kelsall98}. We therefore aim to improve the IPD cloud model using the {\it AKARI} mid-IR all-sky survey data, the latest IR all-sky data.

{\it AKARI} \citep{murakami07}, the Japanese IR astronomical satellite, has two scientific focal-plane instruments, the Infrared Camera (IRC; \citealt{onaka07}) for the wavelength coverage of 1.8--26.5 $\mu$m and the Far-Infrared Surveyor (FIS; \citealt{kawada07}) for 50--180 $\mu$m. {\it AKARI} was launched on 2006 February 21 and was brought into a sun-synchronous polar orbit at an altitude of 700 km. {\it AKARI} started the performance verification operation from 2006 April 24 and carried out all-sky surveys during a period from 2006 May 8 to 2007 August 28 with the telescope cooled at 6 K by liquid helium and mechanical coolers \citep{kaneda05,kaneda07}. The 9 and 18 $\mu$m photometric bands of the IRC and the 65, 90, 140, and 160 $\mu$m bands of the FIS were used for the all-sky surveys.

Among them, we use the IRC mid-IR data (9 and 18 $\micron$ bands) for modeling the zodiacal emission. The {\it AKARI} mid-IR all-sky data enable accurate modeling of the zodiacal emission thanks to higher spatial resolution than the {\it COBE}/DIRBE and the {\it IRAS} data. Moreover, the 9 $\mu$m map is crucial to investigate the all-sky distribution of polycyclic aromatic hydrocarbons, while the 18 $\mu$m map is useful to trace hot dust grains. Therefore, it is also important to model the zodiacal emission accurately for studies of the Galactic interstellar medium (ISM).

\section{Observations and data reduction}
\label{obs}

\subsection{The IRC all-sky survey}
\label{IRC}

The mid-IR all-sky survey was conducted with the two photometric broad band filters centered at 9 $\mu$m ({\it S9W}, effective bandwidth: 4.10 $\mu$m) and 18 $\mu$m ({\it L18W}, effective bandwidth: 9.97 $\mu$m) of the MIR-S and MIR-L channels, respectively \citep{onaka07,ishihara10}. The spectral response curves of these bands are shown in Figure 1 in \citet{ishihara10}. {\it AKARI} revolved around the Earth in a sun-synchronous polar orbit. In the all-sky survey, the satellite scanned the sky along the circle of the solar elongation at approximately 90$^\circ$. The MIR-S and MIR-L channels have 256$\times$256 pixels with the pixel scales of $2\mbox{$.\!\!^{\prime\prime}$}34 \times 2\mbox{$.\!\!^{\prime\prime}$}34$ and $2\mbox{$.\!\!^{\prime\prime}$}51 \times 2\mbox{$.\!\!^{\prime\prime}$}39$, respectively. The full widths at half maxima of the point spread functions for the 9 and 18 $\mu$m bands are $5\mbox{$.\!\!^{\prime\prime}$}5$ and $5\mbox{$.\!\!^{\prime\prime}$}7$, respectively \citep{onaka07,ishihara10}. Two detector rows out of 256 in the sensor array were used for the all-sky observations, and the two rows with the width of 10$\arcmin$ scanned the sky in a continuous and non destructive readout mode \citep{ishihara06,ishihara10}. The interval between scans is 100 min. The orbit rotated about the axis of the Earth in one year, and hence the satellite covered the whole sky in half a year. {\it AKARI} observed the leading and the trailing directions of the Earth orbit alternately for every scan, because the solar elongation was fixed at 90$^\circ$. The zero level of the surface brightness observed by {\it AKARI} is well calibrated for diffuse sources because of accurate dark current measurement. The dark current was measured with the cold shutter closed during the maneuver operation of the satellite. The dark current measurement took $\sim$10 min and was performed about seven times per day on average. We evaluate the average and standard deviation of the dark level using all the measurements made during the mission period. The dark levels are $1.68\pm 0.03$ MJy~sr$^{-1}$ and $3.87\pm 0.06$ MJy~sr$^{-1}$ at 9 and 18 $\micron$, respectively, and these values are subtracted from the data.

\subsection{Production of the all-sky maps}
\label{production}

First, we carried out corrections for reset anomaly and non-linearity of the detector \citep{ishihara10}. Secondly, we corrected the effects of ionizing radiation by cosmic rays \citep{mouri11} and scattered light from the moon. Finally, we converted the unit of the surface brightness from an analog-to-digital unit (ADU) to MJy~sr$^{-1}$, using the following conversion factors which are described in the README document of the {\it AKARI} mid-IR all-sky diffuse maps prepared for the public data release: {\it S9W}: 1 ADU$=$0.303 MJy~sr$^{-1}$ and {\it L18W}: 1 ADU$=$0.474 MJy~sr$^{-1}$.

\begin{figure}
\plotone{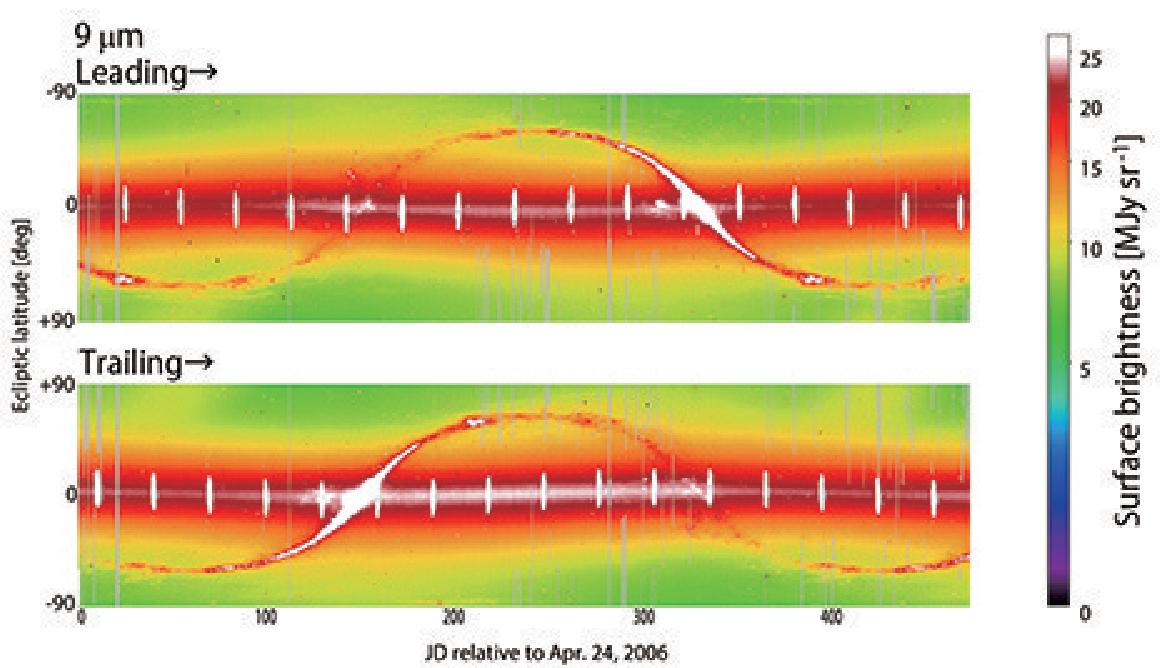}
\plotone{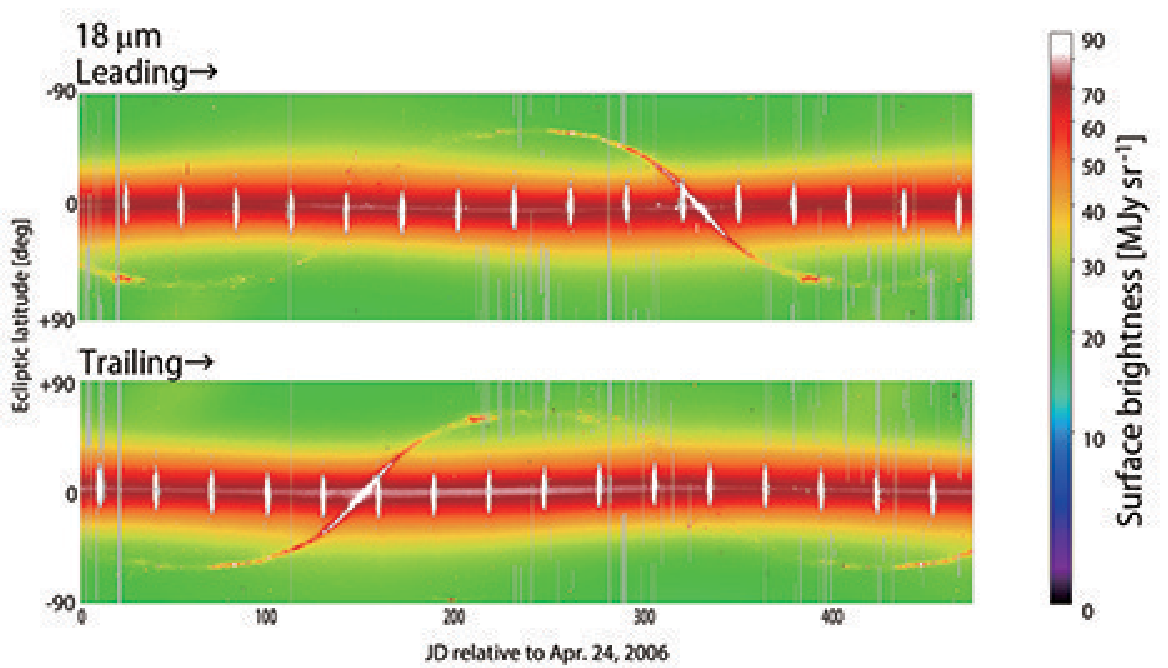}
\caption{{\it AKARI} mid-IR all-sky diffuse maps on the plane of the ecliptic latitude versus the Julian day relative to 2006 April 24, where the scan direction is from bottom to top. For each of the 9 and 18 $\micron$ maps, the upper and the lower panels correspond to the leading and the trailing sides, respectively. The arrow shown in the top left of each map indicates the direction of the shift of the scan path, which is useful for comparison with the DIRBE maps in Figure \ref{fig:dirbe}.}
\label{fig:allsky_day}
\end{figure}

With these processes, we obtained the {\it AKARI} mid-IR all-sky diffuse maps as shown in Figure \ref{fig:allsky_day}. In this figure, all the scan data are plotted. The horizontal and the vertical axes indicate the observation days and the ecliptic latitude, respectively. The scan direction is from bottom to top in this figure.

\section{Modeling of the zodiacal emission}
\label{model}

The zodiacal emission brightness varies with time because of the orbital motion of the Earth relative to the IPD cloud, while the brightnesses of the other components such as the Galactic component do not vary. For determination of the model parameters, \citet{kelsall98} used the seasonal variation of the brightness in the DIRBE maps as a time-varying component of the zodiacal emission brightness, in order to separate the zodiacal emission from the other components. However, we consider that the modeling which considers only time-varying components cannot reproduce the overall distribution of the zodiacal emission very well, because there remains the residual component, whose level is higher than 0.5 MJy~sr$^{-1}$, in the DIRBE 12 and 25 $\mu$m maps \citep{kelsall98} and the {\it AKARI} 9 $\mu$m map \citep{pyo10}. We therefore introduce a new method; we determine the parameters in the Kelsall model using the absolute brightness (i.e., both time-varying and non-time-varying components) of the zodiacal emission in the {\it AKARI} mid-IR maps.

\subsection{The Kelsall model}
\label{kelsall_model}

Although the detail of the Kelsall model is described in \citet{kelsall98}, we give basic information here. In the Kelsall model, the zodiacal emission is decomposed into the following three components: a smooth cloud, dust bands, and a MMR component. The smooth cloud is a dominant component, which has been studied for a historically long time by ground-based observations (e.g., \citealt{cassini1685}). The origin of this component is considered to be a mixture of dust associated with asteroidal and cometary debris. The dust bands represent asteroidal collisional debris \citep{dermott84,nesvorny06,nesvorny08}, which were first discovered by {\it IRAS} \citep{low84}. 
The IPD grains are slowly spiraling inward by the Poynting-Robertson effect and the dust bands are thought to be produced by replenishment from main-belt asteroids. 
The Kelsall model contains three pairs of the dust bands which appear at the ecliptic latitudes around $\pm 1\mbox{$.\!\!^{\circ}$}4$, $\pm 10^\circ$, and $\pm 15^\circ$, based on the structure revealed by \citet{reach97} with the spatially-filtered DIRBE and {\it IRAS} data. The MMR component is composed of migrating dust grains, which are temporarily trapped into resonant orbits near 1 AU by the Earth. This component had been expected theoretically \citep{jackson89,marzari94a,marzari94b,dermott94,dermott96}, and the existence was confirmed by \citet{reach95} based on the DIRBE data. The MMR component has an asymmetric structure, which is decomposed into a circumsolar ring and a three-dimensional Earth-trailing blob.

The brightness of the zodiacal emission observed at the wavelength $\lambda$ (i.e., 9 or 18 $\mu$m), at the celestial position $p$, and at the time $t$ is calculated by integrating the thermal emission along the line of sight $s$, summed for the zodiacal emission components $c$ as
\begin{equation}
	S_{\lambda}(p,t)=\sum_{c=1,2,3} \int n_{c}(x,y,z)\varepsilon_{\lambda}B_{\lambda}(T)ds,
	\label{zodi_brightness}
\end{equation}
where $n_{c}(x,y,z)$ is the density of each component at position $(x,y,z)$ and $\varepsilon_{\lambda}$ is the emissivity modification factor that measures deviations from the Planck function $B_{\lambda}(T)$. For simplicity, in this study, we use the same values of $\varepsilon_{\lambda}$ for the above three components, although there are suggestions that the far-IR emissivity may be different between the components \citep{planck14}. 
We assume that the dust temperature, $T$, varies with the distance from the Sun, $R$, as
\begin{equation}
	T(R)=T_{0}R^{-\delta},
	\label{temperature}
\end{equation}
where $T_0$ is 286 K, a dust temperature at 1 AU and $R$ is given in units of AU. We normalize $\varepsilon_{\lambda}$ to unity at 18 $\mu$m, while \citet{kelsall98} normalized $\varepsilon_{\lambda}$ to unity at 25 $\mu$m.

\subsection{Fitting technique}
\label{tech}

\begin{figure}
\plotone{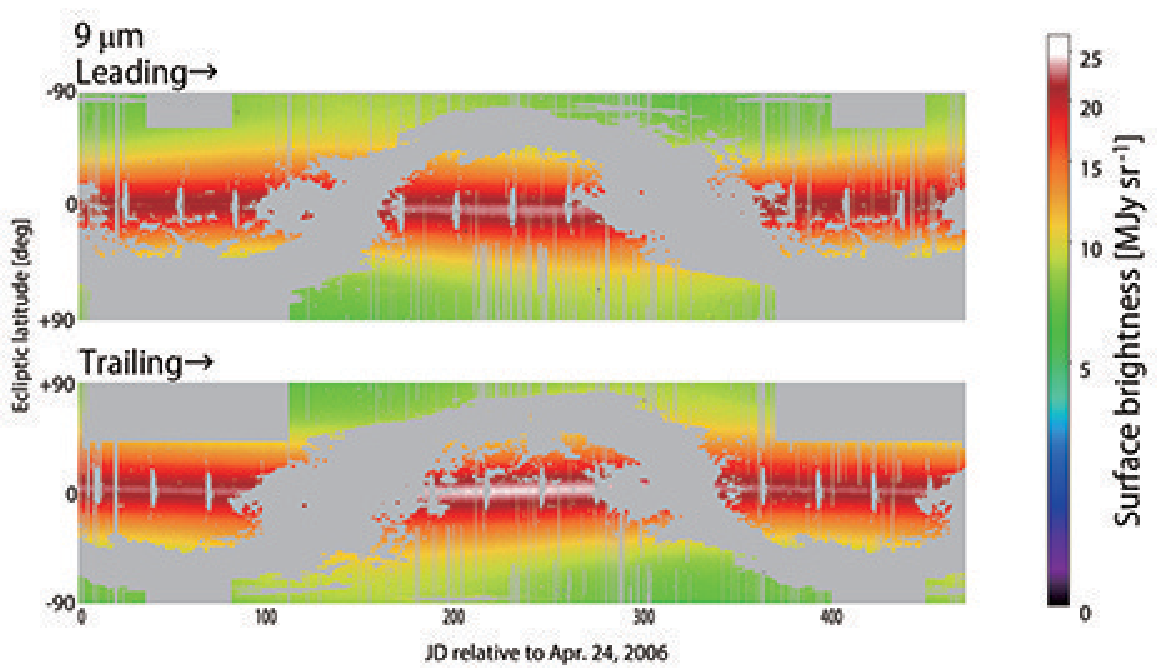}
\plotone{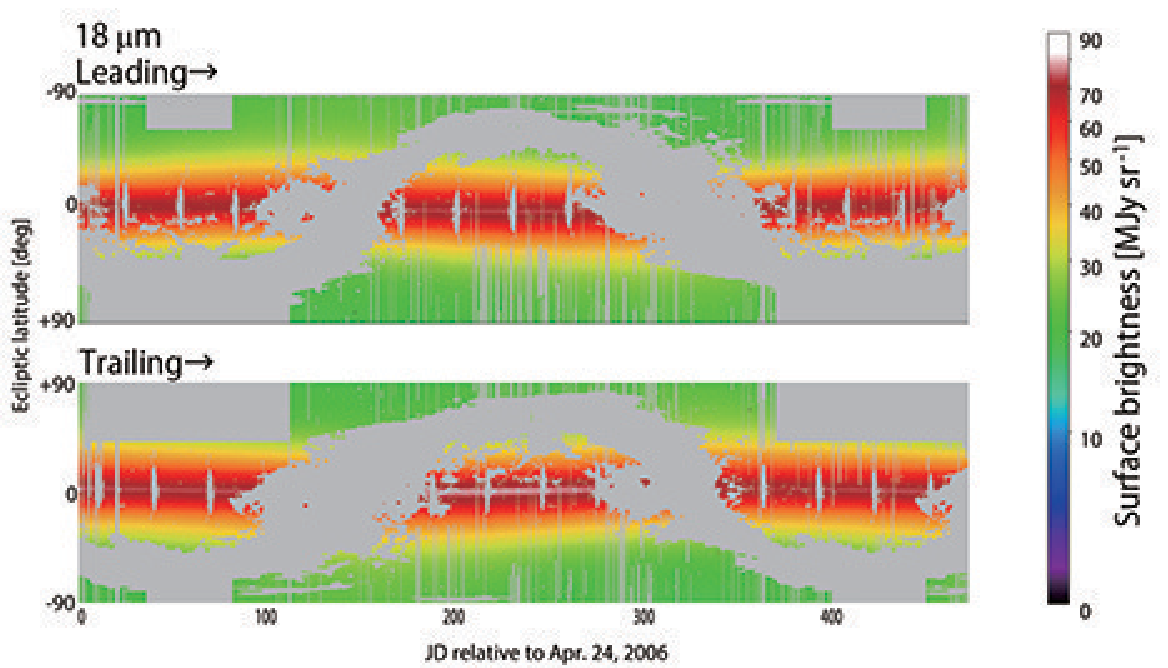}
\caption{Same as Figure \ref{fig:allsky_day}, but after masking the regions affected by the Galactic emission, the earthshine, and the scattered light from the moon.}
\label{fig:allsky_mask}
\end{figure}

We carry out model fitting using the absolute brightness of the zodiacal emission in the {\it AKARI} 9 and 18 $\mu$m all-sky maps. We determine the model parameters to minimize the $\chi^2$ value. We use the all-sky maps in Figure \ref{fig:allsky_day} with spatial sampling of 10$\arcmin$ for the fitting. In order to avoid the Galactic component, we masked the regions of surface brightness levels higher than 6 MJy~sr$^{-1}$ in the {\it AKARI} 140 $\micron$ data \citep{doi15}. Considering typical interstellar colors of the mid- to far-IR brightness, the 6 MJy~sr$^{-1}$ limit at 140 $\micron$ corresponds to about 0.1 MJy~sr$^{-1}$ at 9 and 18 $\micron$, which is 2--3 times higher than the brightness fluctuations in the 9 and 18 $\micron$ maps. In addition to the Galactic component, the {\it AKARI} maps suffer stray light from the Earth near the ecliptic poles, with the brightnesses of $\sim$1 and $\sim$5 MJy~sr$^{-1}$ in the 9 and 18 $\mu$m bands, respectively. This stray light corresponds to the earthshine scattered on the telescope baffle only in the summer season \citep{FIS,IRC,pyo10}. We therefore did not use the regions observed in the period of May 5 to August 13 in 2006 and 2007 at ecliptic latitudes higher than 40$^\circ$ and the regions observed in the period of May 25 to July 14 in 2006 and 2007 at ecliptic latitudes lower than $-60^\circ$. We also masked the regions of the lunar elongations lower than 17$\arcdeg$ because the correction for the scattered light from the moon is not sufficient in these regions. Figure \ref{fig:allsky_mask} shows the maps after masking those components.

The Kelsall model intrinsically has 50 model parameters excluding those related to the scattering in the near-IR. Among them, we treat 24 parameters as free parameters (Table \ref{table:1}), which are to be determined by the {\it AKARI} 9 and 18 $\mu$m band data in the present study. We mainly focus on large-scale structures of the IPD cloud in this study, and thus do not determine most of the geometrical parameters of the dust bands. We cannot determine the radial distribution of the MMR component because the solar elongation was fixed at 90$\arcdeg$. The relations between each parameter are complicated because the observed brightness is expressed by integrating the thermal emission along the line of sight and summed for the three cloud components. In order to robustly determine the parameters, we model the zodiacal emission on a large-scale and a small-scale structure separately by taking the following step-by-step procedure. About the brightness uncertainties in the model fitting, we adopt 1 MJy~sr$^{-1}$ considering all the calibration errors associated with the data corrections described in Section \ref{production}.

\subsubsection{One-dimensional fitting at the poles and on the plane}
\label{1D}

\begin{figure}
\plottwo{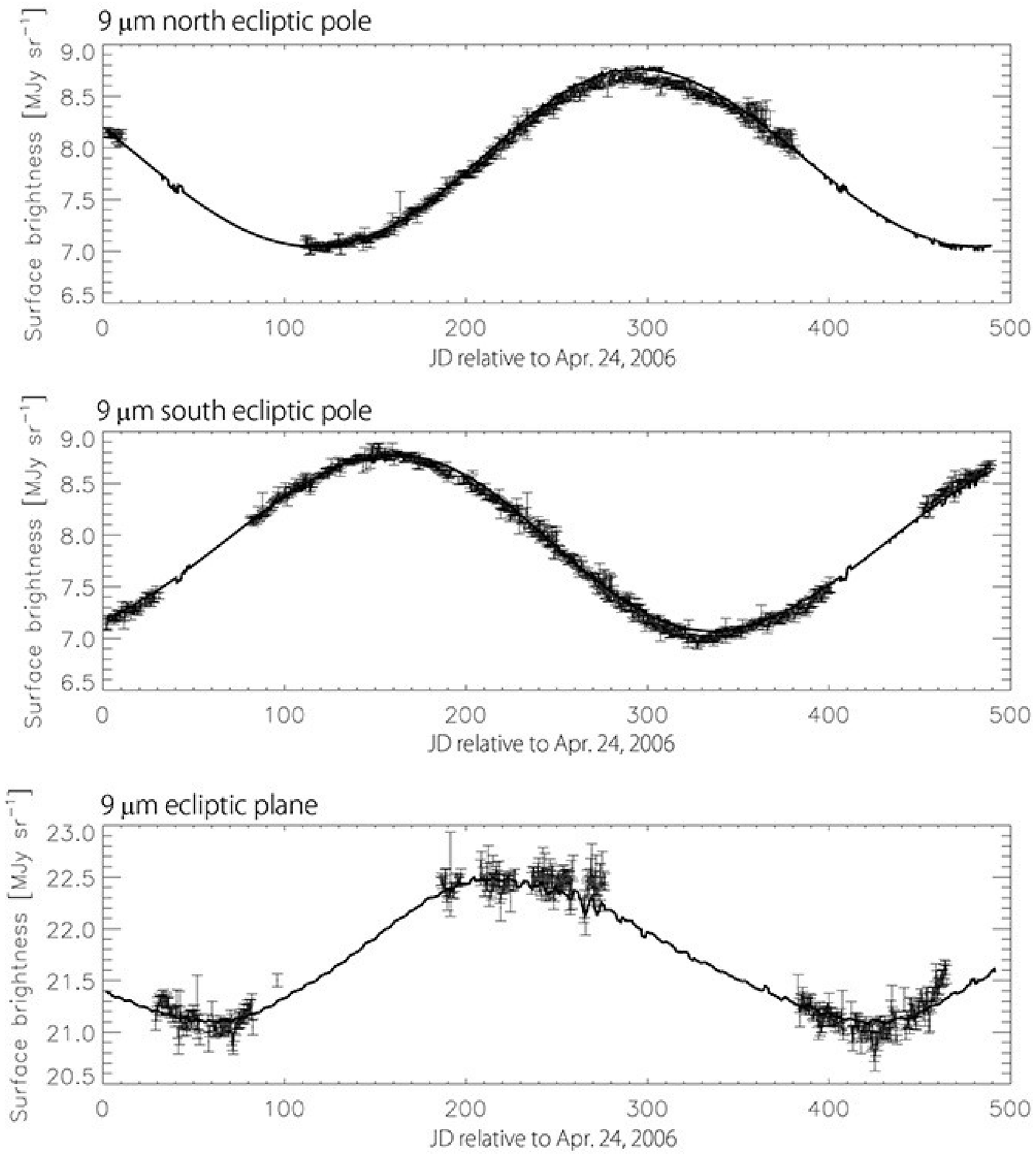}{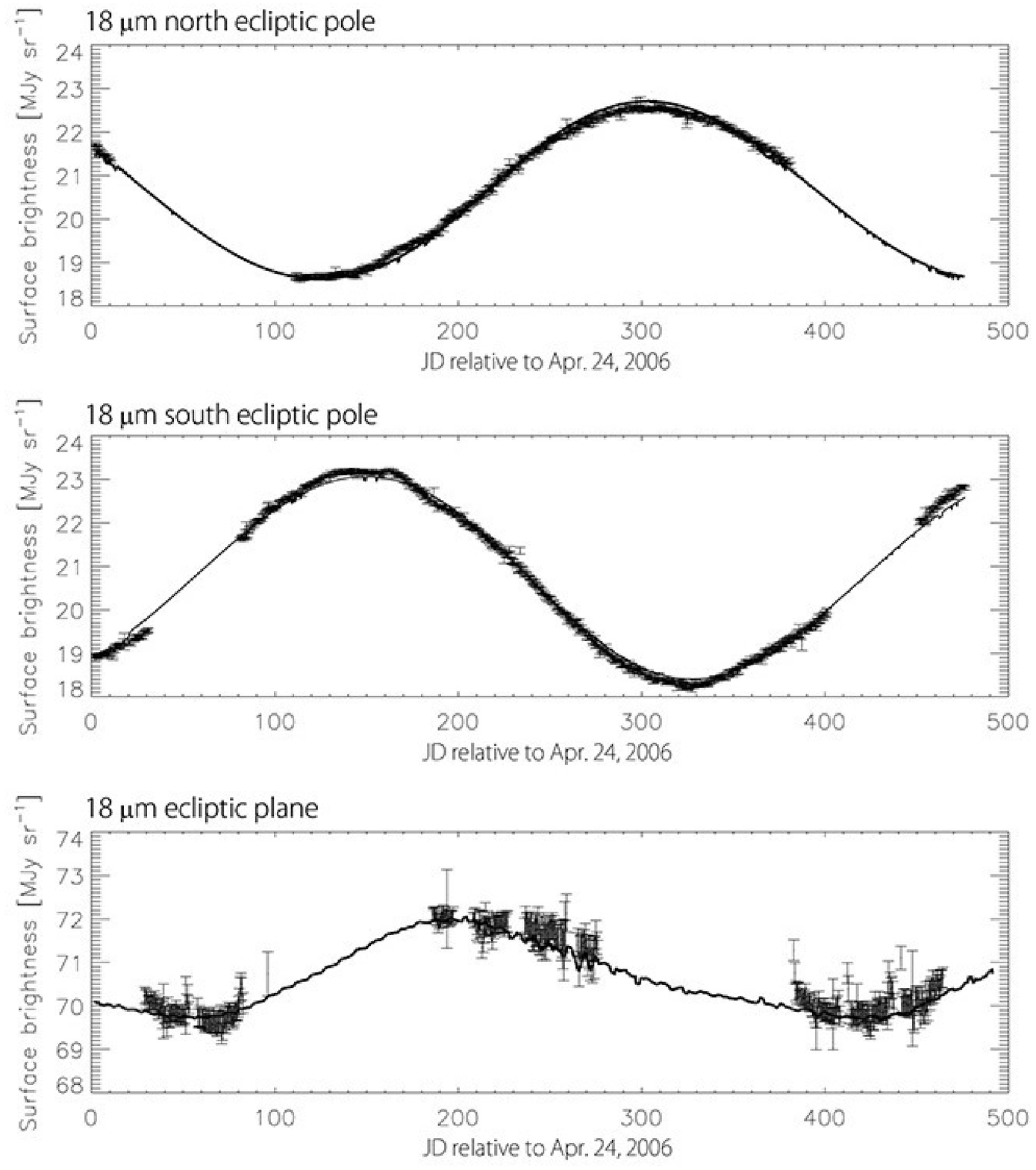}
\caption{Surface brightness profiles observed at the ecliptic poles and on the ecliptic plane, where the error bars indicate the standard deviation of the brightness fluctuations calculated from the surrounding 7$\times$5 grids. Curves represent the best-fit model.}
\label{fig:plot}
\end{figure}

We determined large-scale geometrical parameters of the smooth cloud based on one-dimensional brightness profiles at the ecliptic poles and on the ecliptic plane as shown in Figure \ref{fig:plot}. These profiles were created from the regions of absolute values of ecliptic latitudes higher than 89$^\circ$ for the ecliptic poles and lower than 0$\fdg$4 for the ecliptic plane. We used the data taken on the Earth-leading side, which are not affected by the Earth-trailing blob. Here we temporarily fixed the parameters of smaller-scale structures at those in the Kelsall model, which include the vertical structure of the smooth cloud and the parameters for the dust bands and the circumsolar ring.

First, we fitted the brightness profiles at the ecliptic poles. The brightness variations at the ecliptic poles are caused by the inclination of the smooth cloud with respect to the ecliptic plane and the offset from the Sun of the cloud center. We therefore determined the inclination angle, $i$, the ascending node, $\Omega$, and the offset, $(X_0,Y_0,Z_0)$. We also determined the dust density at 1 AU, $n_0$, and the emissivity modification factor at 9 $\micron$, $\varepsilon _{9\micron}$, from the amplitudes of the variations and ratios of the 9 to 18 $\micron$ brightness, respectively. Next, we fitted the brightness profiles on the ecliptic plane. The brightness variations on the ecliptic plane are caused by the eccentricity of the Earth's orbit, which means that the zodiacal emission brightness is highest at the perihelion and lowest at the aphelion. We determined the radial power-law index of the dust density, $\alpha$, and that of the dust temperature, $\delta$, from the difference between the highest and lowest levels in the 9 and 18 $\micron$ bands. Then, we simultaneously fitted the brightness profiles at the poles and on the plane; here the above best-fit parameters were used as initial values and the fitting range was set to be within $\pm 10\sigma$ where $\sigma$ was the errors estimated by the above fitting.

\subsubsection{Two-dimensional fitting of the all-sky maps}
\label{2D}

Using the two-dimensional maps taken on the Earth-leading side, we determined the parameters of smaller-scale structures along the scan direction. For the large-scale geometrical parameters of the smooth cloud, the best-fit parameters in the one-dimensional fitting were used as initial values and the fitting range was again set to be within $\pm 10\sigma$. Once we obtained the parameters for the leading-side maps, we subtracted the zodiacal emission with the best-fit model parameters from the trailing-side maps to determine the distribution of the Earth-trailing blob in the 9 and 18 $\micron$ bands. We then determined the parameters of the Earth-trailing blob from the subtracted maps. Finally, we repeated the one-dimensional fitting where we fixed the parameters of the small-scale structures at the best-fit values in the present two-dimensional fitting but not at those in the Kelsall model. Table \ref{table:1} summarizes the best-fit parameters thus obtained.

\section{Results}
\label{results}

\begin{figure}
\plotone{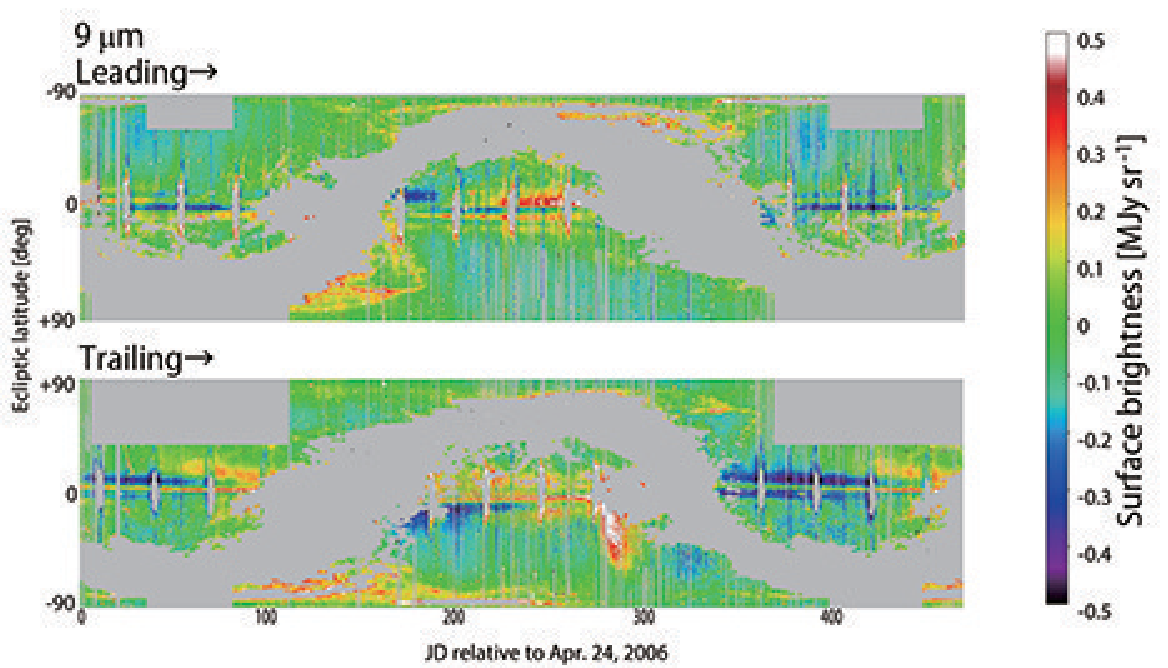}
\plotone{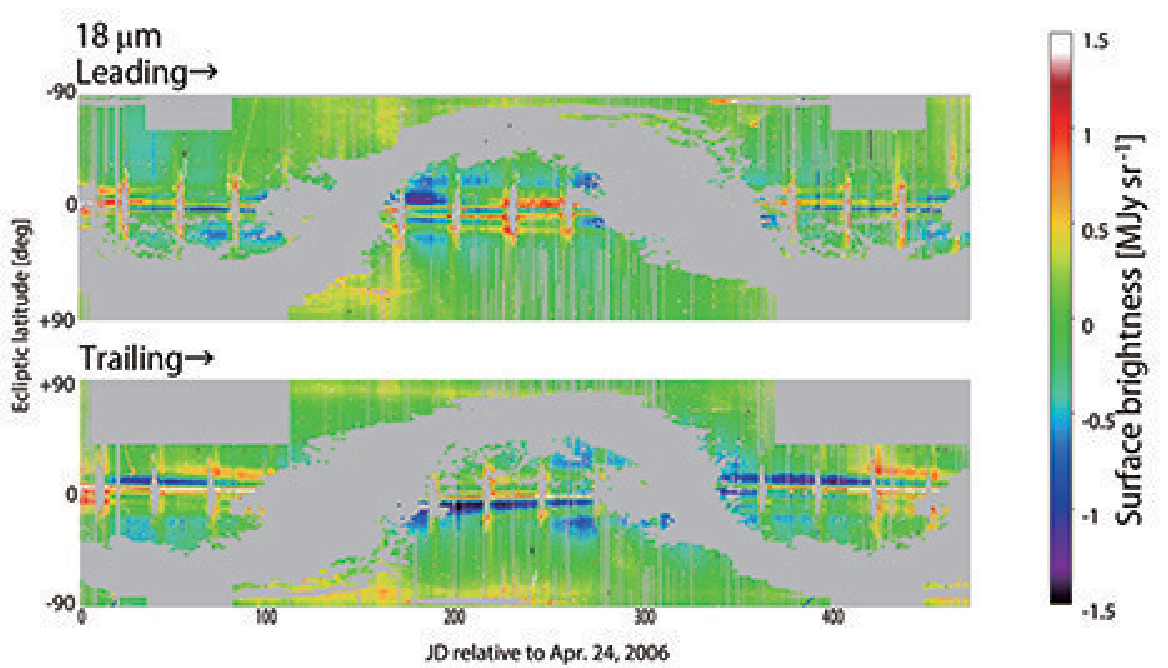}
\caption{{\it AKARI} mid-IR maps after subtraction of the zodiacal emission with our new model. Note that the color scales are different from Figure \ref{fig:allsky_mask} to emphasize the faint residual components.}
\label{fig:allsky_rmn}
\end{figure}

\begin{figure}
\plotone{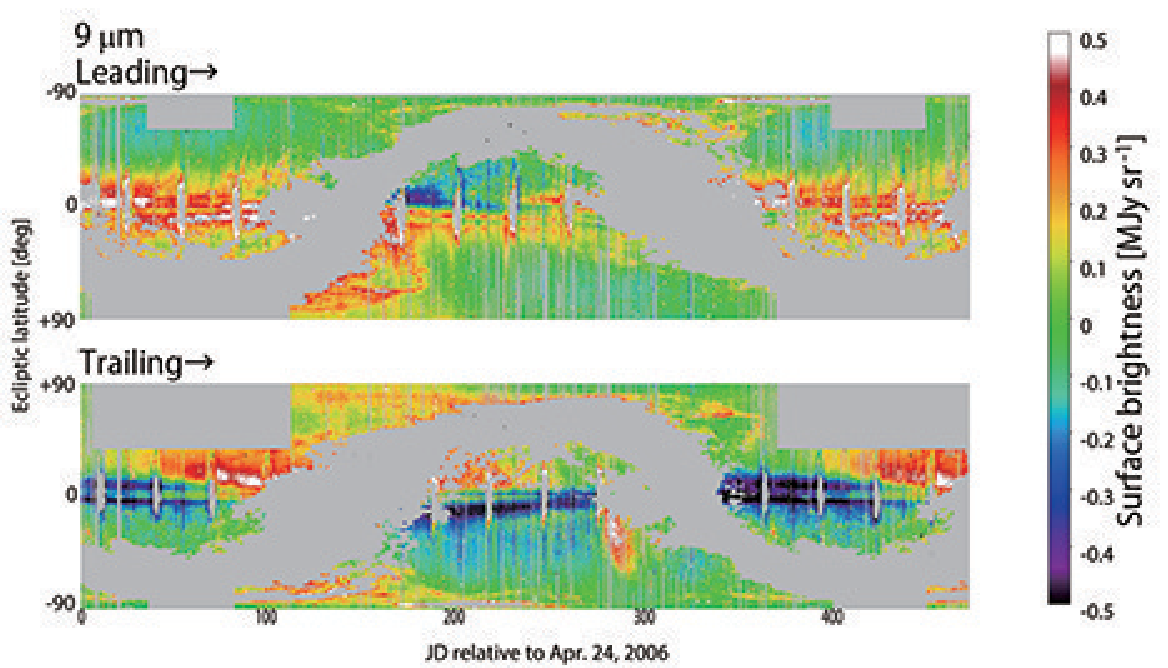}
\plotone{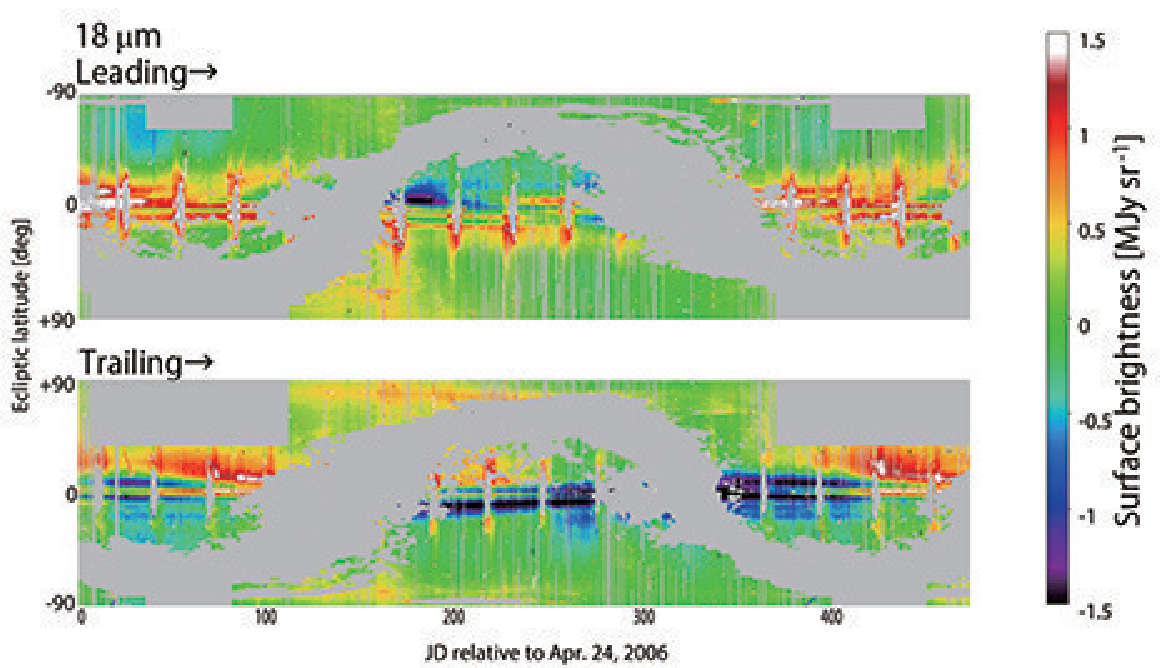}
\caption{Same as Figure \ref{fig:allsky_rmn}, but the Kelsall model is used for subtraction of the zodiacal emission.}
\label{fig:allsky_kelsall}
\end{figure}

Figure \ref{fig:allsky_rmn} shows the {\it AKARI} mid-IR all-sky maps after subtraction of the zodiacal emission obtained with our new model. For comparison, the residual maps after subtraction of the zodiacal emission with the Kelsall model are also shown in Figure \ref{fig:allsky_kelsall}, where the emissivity modification factors and the constant component levels at 9 and 18 $\mu$m are determined by the {\it AKARI} data. As can be seen in these maps, our new model successfully improves the residual levels and the gradient of the residuals along the observation days. The root mean square (RMS) values of the residual components improve from 0.16 to 0.08 MJy~sr$^{-1}$ at 9 $\micron$ and from 0.31 to 0.20 MJy~sr$^{-1}$ at 18 $\mu$m from the Kelsall model to our model.

The model parameters thus optimized are summarized in Table \ref{table:1} in comparison with the result in \citet{kelsall98}. The power-law index of the radial distribution of the dust density $\alpha$ and the vertical shape parameter $\beta$ of the smooth cloud become significantly larger than those of the Kelsall model. This implies that the size of the smooth cloud is by about 10\% more compact than previously thought. For the dust bands, since we did not change most of the geometrical parameters, only the density parameters are listed in Table \ref{table:1}. However, in the {\it AKARI} mid-IR maps, we cannot recognize the narrow double peak structure along the ecliptic latitude of band 2, the strongest dust band, which are adopted in the Kelsall model. We have therefore removed the contribution of the double peak structure of band 2 in the model. In addition, we find that the Earth-trailing blob is likely to have a vertical offset, $Z_{\rm 0,TB}$, of $0.0206\pm 0.0005$ AU to the north direction.

\begin{figure}
\plotone{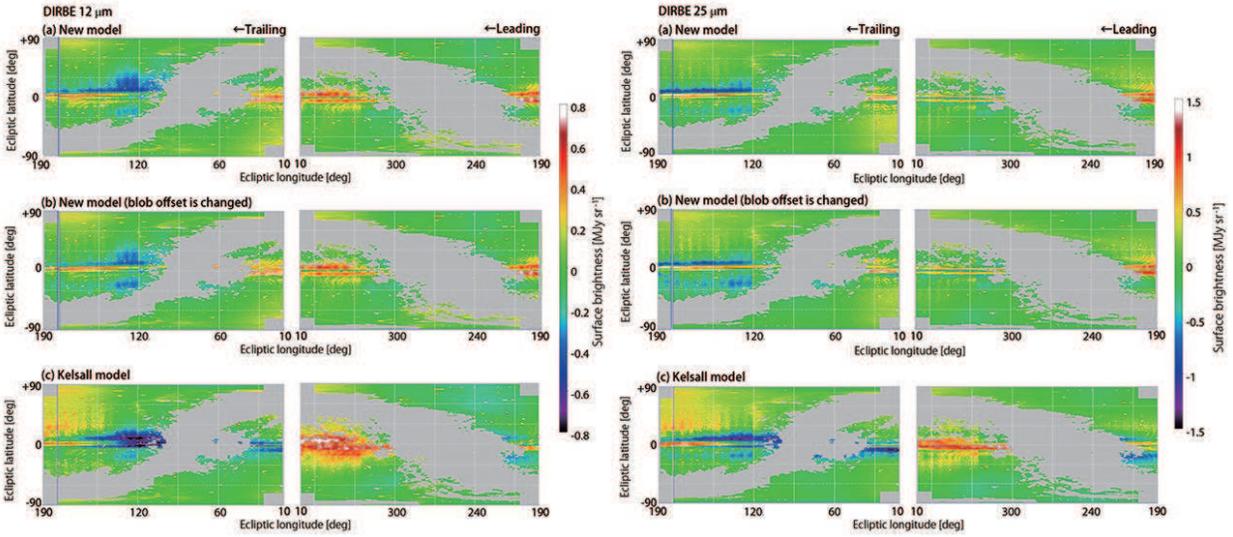}
\caption{DIRBE Solar Elongation $=$ 90 deg Sky Maps in the 12 and 25 $\mu$m bands in the ecliptic coordinates, after subtraction of the zodiacal emission with the models denoted in the figure. (a) Residuals after subtraction of the zodiacal emission with our new model. (b) Residuals after subtraction of the zodiacal emission with our new model where the vertical offset of the Earth-trailing blob, $Z_{\rm 0,TB}$, is changed to better fit the DIRBE data. (c) Residuals after subtraction of the zodiacal emission with the Kelsall model. The arrow shown in the top right of each map indicates the direction of the shift of the scan path, which is useful for comparison with the {\it AKARI} maps in Figure \ref{fig:allsky_rmn}.}
\label{fig:dirbe}
\end{figure}

In order to confirm whether these changes are due to model improvements or temporal variations of the zodiacal emission, we apply our new model to the DIRBE data. Figure \ref{fig:dirbe}(a) shows the DIRBE Solar Elongation $=$ 90 deg Sky Maps in the 12 and 25 $\mu$m bands in the ecliptic coordinates, after subtraction of the zodiacal emission with our new model. These maps contain both leading and trailing sides obtained in half a year observations; the region at ecliptic longitudes from 10$^\circ$ to 190$^\circ$ corresponds to the trailing side, while the other region corresponds to the leading side. For comparison with the {\it AKARI} maps, the ecliptic longitudes in these maps are equivalent to the observation days and the direction of the shift of the scan path is indicated by the arrows in Figure \ref{fig:dirbe}.

As can be seen in this figure, the model over-predicts the brightness in the north regions on the trailing side. The regions on the trailing side in these maps correspond to the regions with Day 250 to 440 in the {\it AKARI} maps in Figure \ref{fig:allsky_rmn}, where the residual pattern is not so asymmetrical with respect to the ecliptic plane as the DIRBE map. Hence the Earth-trailing blob component in our new model does not fit the DIRBE data well, suggesting that the geometry of the blob component has changed from the {\it COBE} epoch to the {\it AKARI} epoch. Since the over-subtracted region is systematically shifted to the north with respect to the ecliptic plane, the vertical offset, $Z_{\rm 0,TB}$, should be smaller for the DIRBE data than for the {\it AKARI} data. In order to reduce the asymmetry, we re-determine $Z_{\rm 0,TB}$ for the DIRBE data using the 12 $\micron$ map because the asymmetry is clearer in the 12 $\micron$ map. The best-fit $Z_{\rm 0,TB}$ for the DIRBE data is $0.007\pm 0.002$ AU, which is smaller than that for the {\it AKARI} data. The change in $Z_{\rm 0,TB}$ is significant (7$\sigma$), based on the $\chi ^2$ statistics. Then, we apply this value to the model for the DIRBE data as shown in Figure \ref{fig:dirbe}(b). For comparison, we also show the residual maps after subtraction of the zodiacal emission with the Kelsall model in Figure \ref{fig:dirbe}(c). Comparing Figure \ref{fig:dirbe}(b) with \ref{fig:dirbe}(c), we confirm that the DIRBE data are better fitted by our new model than by the Kelsall model. The RMS values of the residual components improve from 0.20 to 0.12 MJy~sr$^{-1}$ at 12 $\micron$ and from 0.31 to 0.20 MJy~sr$^{-1}$ at 25 $\mu$m from the Kelsall model to our model. Hence, we conclude that the changes of the model are mostly due to model improvements, except for the vertical offset of the Earth-trailing blob.

\section{Discussion}
\label{discussion}

\subsection{Properties of the IPD cloud}
\label{pro_IPD}

Comparing the parameters of our new model with those of the Kelsall model, we discuss the physical properties of the IPD cloud. The power-law index of the radial distribution of the dust density of the smooth cloud, $\alpha$, becomes significantly larger than that of the Kelsall model from 1.34 to 1.59, which suggests that the cloud size has shrunk in our new model. The radial distribution of the dust density depends on the orbital evolution of the IPD grains. If the orbital evolution of the IPD grains is dominated by the Poynting-Robertson effect, the equilibrium distribution of the dust is expected to have $\alpha =1$ \citep{burns79}. However, not only the present study but also \citet{kelsall98} and other previous studies show the values significantly larger than the expected one. \citet{dumont75} obtained $\alpha =1.2$ by the ground-based observations and \citet{leinert81} obtained $\alpha =1.3$ for $R\leq 1$ AU by space observations with the {\it Helios 1} and {\it 2} space probes. Among them, the present result shows the largest difference from the expected value. To interpret the difference between the expected and the observed values, the previous studies have suggested dust supply mechanisms in inner regions other than migrating dust from the outer Solar System. For example, \citet{leinert83} and \citet{grun85} showed that the probability of the dust production due to the collision of large grains increases with the decreasing distance from the Sun, and \citet{ishimoto2000} suggested that a large amount of the dust may be supplied by comets around 1 AU. Our result strongly supports the necessity of such dust supply around 1 AU.

On the other hand, the power-law index of the radial distribution of the dust temperature, $\delta$, is not significantly changed from the Kelsall model. Both $\delta$ values in our new model and the Kelsall model are significantly different from 0.5 which is expected if the dust grains emit the blackbody radiation in the thermal equilibrium. Our result therefore indicates that the dust sizes are not large enough to emit blackbody radiation in the mid-IR.

We find that the Earth-trailing blob has moved to the north direction by about 0.014 AU between the {\it COBE} and the {\it AKARI} epoch. It may suggest that the dust is recently supplied by comets near the Earth, asymmetrically trapped on the trailing side of the Earth (i.e., a larger amount or closer to the Earth in the north than in the south).

\subsection{The constant component}
\label{pro_const}

\begin{figure}
\plottwo{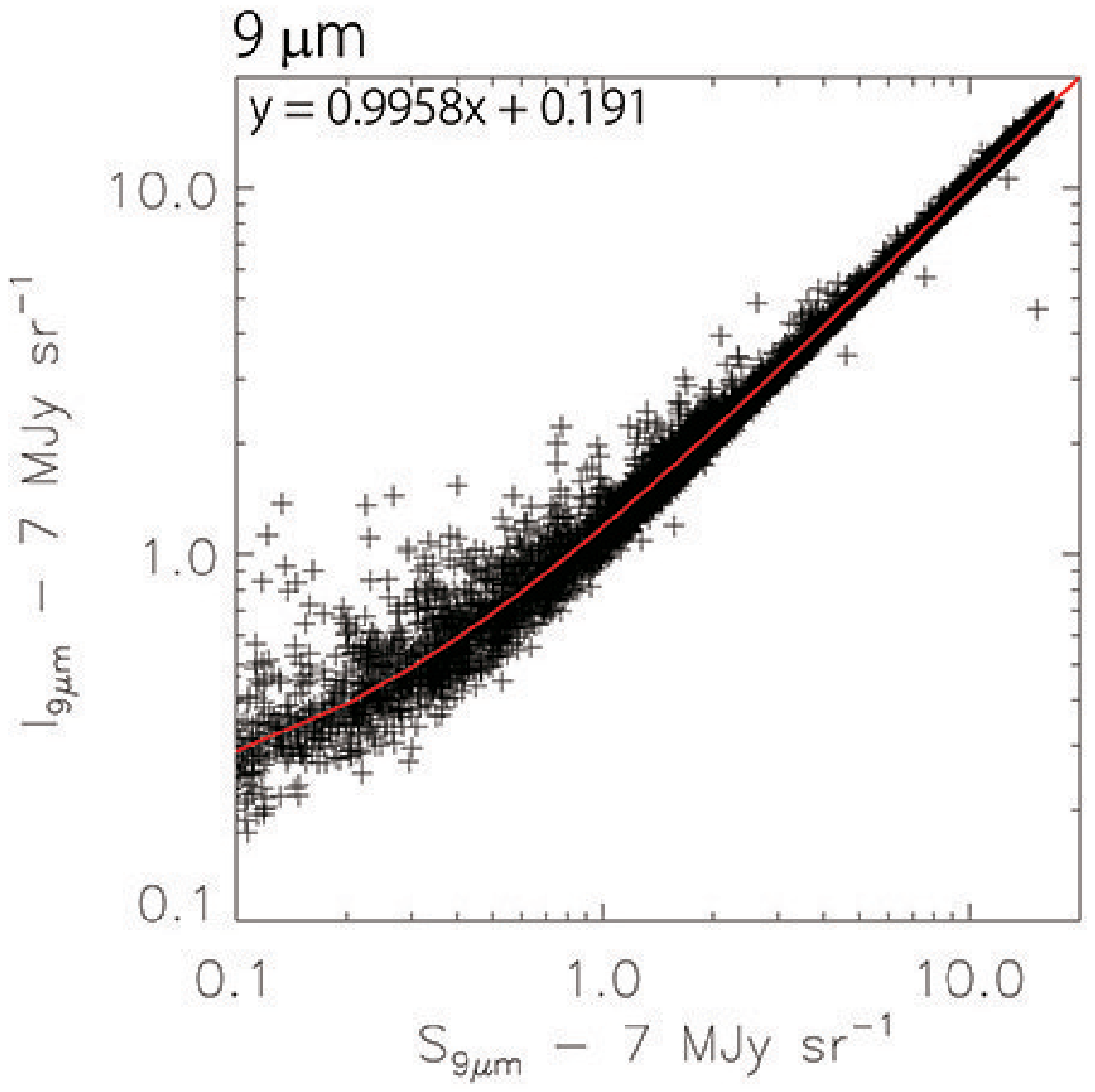}{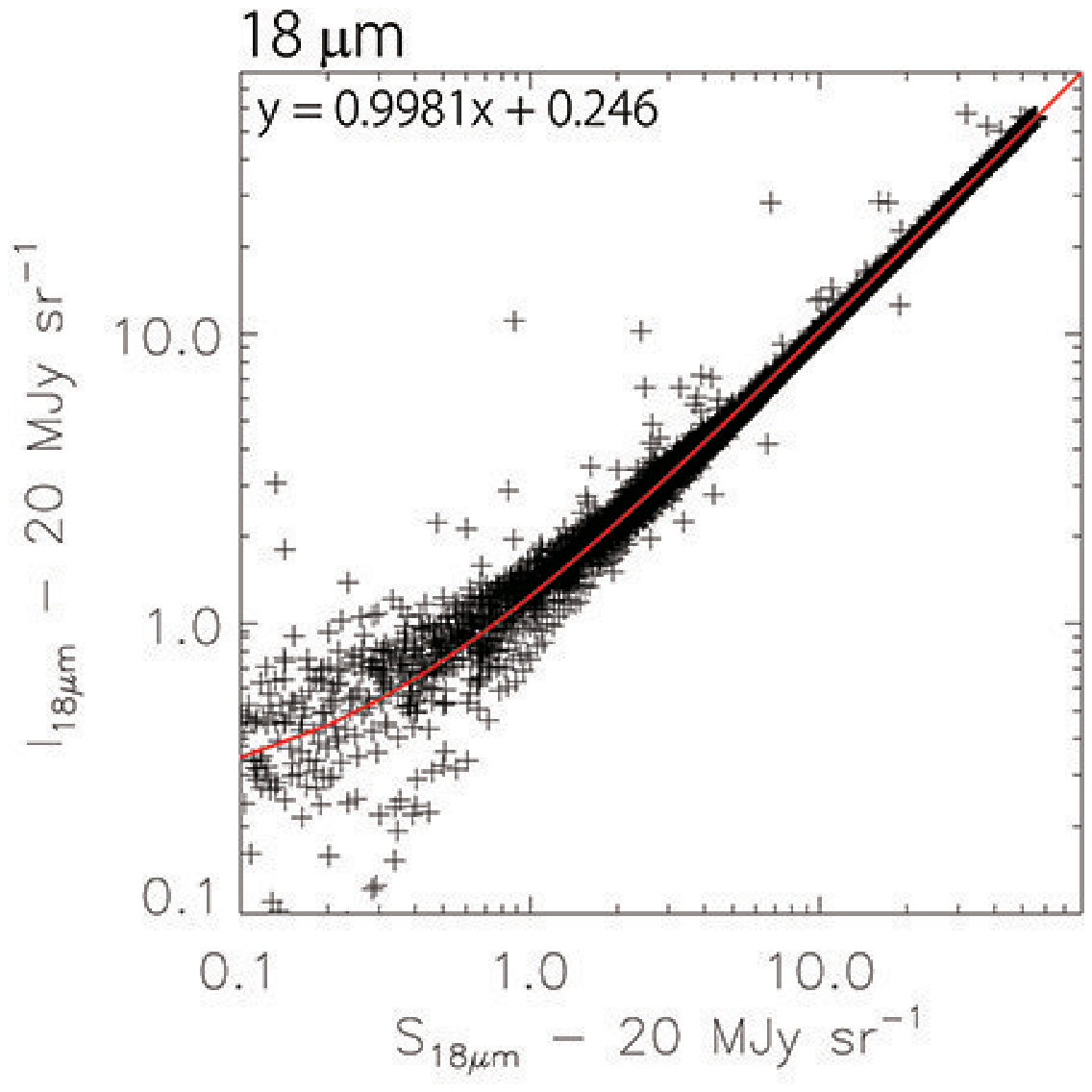}
\caption{Correlation plots between the observed brightness $I_\lambda$ and the predicted brightness $S_\lambda$ by the model excluding the constant component. In order to show the presence of a small offset in the observed brightness, we subtract 7 and 20 MJy~sr$^{-1}$ at 9 and 18 $\mu$m, respectively, from both $I_\lambda$ and $S_\lambda$ values for a display purpose. Lines represent the best-fit result of linear fitting to the data points.}
\label{fig:soukan}
\end{figure}

In order to evaluate the brightness of the constant component, $C_\lambda$, we fit the relation between the observed brightness and the brightness predicted by the model excluding the constant component with a line as shown in Figure \ref{fig:soukan}. The intercepts of the fitted lines are $0.191\pm 0.004$ MJy~sr$^{-1}$ and $0.246\pm0.009$ MJy~sr$^{-1}$ at 9 and 18 $\micron$, respectively, while the slopes are $0.9958\pm 0.0003$ and $0.9981\pm 0.0002$ at 9 and 18 $\micron$, respectively. 
Thus even considering the 3$\sigma$ error of the dark current measurement mentioned in Section \ref{IRC}, the presence of the constant component is significant. This component can be Galactic diffuse ISM, cosmic IR background (CIB) (e.g., \citealt{hauser98,arendt98}), or IPD grains of interstellar origins (e.g., \citealt{rowan13}). \citet{arendt98} estimated the DIRBE 12 $\micron$ brightness of the Galactic diffuse ISM emission to be $\sim 0.03$ MJy~sr$^{-1}$ in the high ecliptic latitude regions. The CIB model of \citet{fall96} showed that the brightness of CIB is lower than 0.01 MJy~sr$^{-1}$ in the DIRBE 12 $\micron$ band. These values are about two orders of magnitude smaller than the brightness of the constant component evaluated from the {\it AKARI} data. Therefore, emission from the IPD grains of interstellar origins is likely to make a dominant contribution to the constant component.

Since the IPD grains of interstellar origins are considered to have sizes smaller than IPD grains of internal origins (e.g., \citealt{grun93}), we used the modified blackbody model to estimate the temperature of the constant component from the 9 and 18 $\micron$ brightnesses. For the emissivity power-law index of unity, we obtain the temperature of $270\pm 30$ K. This value is reasonable for the temperature of dust grains at $\sim$1 AU.
\\
\\
Finally, as can been seen in Figure \ref{fig:allsky_rmn}, a bright residual component is extended around the location of $\sim$290 Day and $\sim -30^\circ$. This component is recognized only in the trailing-side map at 9 $\micron$ and therefore it is not likely due to the Galactic emission that was not masked. The morphology suggests that it may not originate from the residual of the zodiacal emission subtraction. That large structure cannot be explained by any instrumental artifacts. Thus, we suggest that this may be a small cloud crossing the Earth's orbit. We will discuss this component more carefully in a separate paper.

\section{Summary}
\label{summary}

We have carried out the modeling of the zodiacal emission, using the {\it AKARI} mid-IR all-sky data which have higher spatial resolution than the DIRBE and the {\it IRAS} data. 
In order to robustly determine the model parameters, we model the zodiacal emission on a large-scale and a small-scale structure separately by dividing the procedure into the one-dimensional and the two-dimensional fitting. As a result, we have succeeded in reducing the residual levels after subtraction of the zodiacal emission from the {\it AKARI} data and thus in improving the modeling of the zodiacal emission. We also confirm that our new model better reproduces the zodiacal emission in the DIRBE data than the Kelsall model, except for the Earth-trailing blob component. Hence we conclude that the changes from the Kelsall model to our new model are mostly due to the model improvements but not the temporal variations. Only for the Earth-trailing blob, we find that its vertical offset increases by about 0.014 AU to the north direction from the {\it COBE} epoch to the {\it AKARI} epoch. The model parameters thus obtained indicate that the cloud size of our new model is more compact than that of the Kelsall model, and that the dust sizes are not large enough to emit blackbody radiation in the mid-IR. We evaluate the surface brightness of the constant component as $0.191\pm 0.004$ MJy~sr$^{-1}$ and $0.246\pm 0.009$ MJy~sr$^{-1}$ at 9 and 18 $\micron$, respectively. 
The color temperature of the dust emission is estimated to be $270\pm 30$ K assuming the modified blackbody radiation with the emissivity power-law index of unity.

\acknowledgments

We express many thanks to the referee for giving us many valuable comments. 
We thank all the members of the {\it AKARI} project. {\it AKARI} is JAXA project with the participation of ESA. This research is financially supported by Grants-in-Aid for JSPS Fellows No. 25002536, Scientific Research (C) No. 25400220, and Young Scientists (A) No. 26707008, and the Nagoya University Program for Leading Graduate Schools, ``Leadership Development Program for Space Exploration and Research'', from MEXT.






\bibliographystyle{apj}
\bibliography{refer}

\clearpage



\begin{deluxetable}{l l c c}
\tabletypesize{\scriptsize}
\tablewidth{0pt}
\tablecaption{Parameters of the IPD model in \citet{kelsall98} and in the present study\label{table:1}}
\tablehead{
\colhead{Parameter}           & \colhead{Name of parameter}      &
\colhead{\citealt{kelsall98}}          & \colhead{Present study}
}
\startdata
	\multicolumn{4}{c}{All cloud components} \\
	\hline
	$T_0$ (K) & Temperature at 1 AU & 286 (fixed) & 286 (fixed) \\
	$\delta$ & Temperature power-law index & 0.467 (0.004) & 0.458 (0.006) \\
	$\varepsilon_{9\,\mu{\rm m}}$ & Emissivity modification factor at 9 $\mu$m &\nodata & 0.8575 (0.0007) \\
	\hline
	\multicolumn{4}{c}{Smooth cloud} \\
	\hline
	$n_{0}$ (AU$^{-1}$) & Density at 1 AU & $1.134\times 10^{-7}$ ($0.006\times 10^{-7}$) & $1.578\times 10^{-7}$ ($0.006\times 10^{-7}$) \\
	$\alpha$ & Radial power-law index & 1.34 (0.02) & 1.59 (0.02) \\
	$\beta$ & Vertical shape parameter & 4.14 (0.07) & 4.85 (0.02) \\
	$\gamma$ & Vertical power-law index & 0.94 (0.03) & 1.043 (0.005) \\
    $\mu$ & Widening parameter & 0.19 (0.01) & 0.180 (0.002) \\
	$i$ (deg) & Inclination & 2.03 (0.02) & 2.047 (0.007) \\
    $\Omega$ (deg) & Ascending node & 77.7 (0.6) & 75.9 (0.1) \\
	$X_0$ (AU) & $x$ offset from the Sun & 0.012 (0.001) & 0.0153 (0.0002) \\
	$Y_0$ (AU) & $y$ offset from the Sun & 0.0055 (0.0008) & $-$0.0081 (0.0002) \\
	$Z_0$ (AU) & $z$ offset from the Sun & $-$0.0022 (0.0004) & $-$0.00160 (0.00004) \\
	\hline
	\multicolumn{4}{c}{Dust bands} \\
	\hline
    $n_{B1}$ (AU$^{-1}$) & Density at 3 AU of band 1 & $5.6\times 10^{-10}$ ($0.7\times 10^{-10}$) & $5.6\times 10^{-10}$ ($0.1\times 10^{-10}$) \\
	$n_{B2}$ (AU$^{-1}$) & Density at 3 AU of band 2 & $2.0\times 10^{-9}$ ($0.1\times 10^{-9}$) & $3.70\times 10^{-9}$ ($0.03\times 10^{-9}$) \\
	$n_{B3}$ (AU$^{-1}$) & Density at 3 AU of band 3 & $1.4\times 10^{-10}$ ($0.2\times 10^{-10}$) & $0.06\times 10^{-10}$ ($0.05\times 10^{-10}$) \\
	\hline
	\multicolumn{4}{c}{Circumsolar ring} \\
	\hline
	$n_{\rm SR}$ (AU$^{-1}$) & Density at 1 AU & $1.8\times 10^{-8}$ ($0.1\times 10^{-8}$) & $0.75\times 10^{-8}$ ($0.06\times 10^{-8}$) \\
	$R_{\rm SR}$ (AU) & Radius of peak density & 1.0282 (0.0002) & 1.0282 (fixed) \\
	$\sigma_{\rm rSR}$ (AU) & Radial dispersion & 0.025 (fixed) & 0.025 (fixed) \\
	$\sigma_{\rm zSR}$ (AU) & Vertical dispersion & 0.054 (0.007) & 0.066 (0.002) \\
	\hline
	\multicolumn{4}{c}{Earth-trailing blob} \\
	\hline
	$n_{\rm TB}$ (AU$^{-1}$) & Density at 1 AU & $1.9\times 10^{-8}$ ($0.1\times 10^{-8}$) & $2.08\times 10^{-8}$ ($0.01\times 10^{-8}$) \\
	$R_{\rm TB}$ (AU) & Radius of peak density & 1.06 (0.01) & 1.06 (fixed) \\
	$\sigma_{\rm rTB}$ (AU) & Radial dispersion & 0.10 (0.01) & 0.10 (fixed) \\
	$\sigma_{\rm zTB}$ (AU) & Vertical dispersion & 0.09 (0.01) & 0.151 (0.001) \\
	$\theta_{\rm TB}$ (deg) & Longitude with respect to Earth & $-$10 (fixed) & $-$10 (fixed) \\
	$\sigma_{\rm \theta TB}$ (deg) & Longitude dispersion & 12 (3) & 11.10 (0.06) \\
	$Z_{\rm 0,TB}$ (AU) & $z$ offset from the Sun & 0.0 (fixed) & 0.0206 (0.0005) \\
	\hline
	\multicolumn{4}{c}{Circumsolar ring + Earth-trailing blob} \\
	\hline
	$i_{\rm RB}$ (deg) & Inclination & 0.49 (0.06) & 0.97 (0.03) \\
	$\Omega_{\rm RB}$ (deg) & Ascending node & 22.28 (0.001) & 296 (1) \\
	\hline
	\multicolumn{4}{c}{Constant component} \\
	\hline
	$C_{9\,\mu{\rm m}}$ (MJy~sr$^{-1}$) & Brightness at 9 $\mu$m & \nodata & 0.191 (0.004) \\
	$C_{18\,\mu{\rm m}}$ (MJy~sr$^{-1}$) & Brightness at 18 $\mu$m & \nodata & 0.246 (0.009) \\
\enddata
\tablecomments{For details of the parameters, see \citet{kelsall98}. The values in the parentheses indicate 1$\sigma$ errors.}

\end{deluxetable}

\end{document}